\documentstyle[twoside,fleqn,espcrc2]{article}

\newcommand{\AmS}{{\protect\the\textfont2
  A\kern-.1667em\lower.5ex\hbox{M}\kern-.125emS}}
\def\frac#1#2{ {{#1} \over {#2} }}

\def\beq{\begin{equation}}
\def\eeq{\end{equation}}
\def\bit{\begin{itemize}}
\def\eit{\end{itemize}}
\newcommand{\ba}         {\begin{eqnarray}}
\newcommand{\ea}         {\end  {eqnarray}}
\newcommand{\ban}        {\begin{eqnarray*}}
\newcommand{\ean}        {\end  {eqnarray*}}

\def\L{\Lambda}
\def\be{\beta}
\def\b{\beta}

\def\eps{\epsilon}

\def\MSbar{\overline{\rm MS}}

\def\np#1#2#3{Nucl.\ Phys.\ B#1 (19#3) #2}

\def\pr#1#2#3{Phys.\ Rev.\ D #1 (19#3) #2}

\title{New issues for Numerical Stochastic Perturbation Theory}

\author{G. Burgio\address{Dipartimento di Fisica, Universit\`a di Parma 
	and INFN, Gruppo Collegato di Parma, Italy}, 
	F. Di Renzo\address{Department of Mathematical Sciences,
	University of Liverpool, United Kingdom}, 
	G. Marchesini\address{Dipartimento di Fisica, Universit\`a di Milano
	and INFN, Sezione di Milano, Italy}, 
        E. Onofri$^{\rm a}$, 
	M. Pepe$^{\rm c}$, 
	and L. Scorzato\thanks{Presenter of the poster.}$^{\rm a}$}

\begin{document}

\begin{abstract}
First attempts in the application of Numerical Stochastic Perturbation
Theory (NSPT) to  the problem of pushing one loop further the computation of
SU(3) (SU(2)) pertubative $\b$ function (in different schemes) are 
reviewed and the relevance of such a computation is discussed.
Other issues include the proposal of a different strategy for gauge-fixed
NSPT computations in lattice QCD. 
\end{abstract}

\maketitle

\section{Perturbative $\beta$-functions by NSPT.}

Perturbative calculations in lattice field theories are 
necessary to connect results  in different renormalisation schemes.
In asymptotically free theories particulary useful are matchings to 
those schemes in which perturbative computations at high energies are
more  easily achieved (first of all $\MSbar$). It is well known that 
the bare definition of the coupling in lattice QCD is often not the 
best one: other definitions are useful, for instance, to relate the 
perturbative and the non perturbative regions in a smoother way.

In the case of QCD with no (or massless) quarks, or in any other pure
gauge theory, just one free parameter is present and every coupling
constant (in any scheme)  can be expanded as a power series in any
other. The first coefficient of such  an expansion determines the
relation between the physical  scales ($\L$ parameter), the
others are related to combinations of  the respective (non universal)
coefficients of the $\beta$-function. 

Aim of the project is to obtain 
higher order expansions of convenient couplings in the lattice scheme, thus 
obtaining informations on higher order $\beta$-function coefficients. The 
natural goal is $\b_3$ in the lattice scheme, which would result for 
example in a more precise correction to asymptotic scaling and in 
pushing one loop further the matching to $\MSbar$.

\section{NSPT for lattice QCD}

Unfortunately perturbation theory presents many difficulties in the lattice
regularisation. 
To bypass such difficulties  NSPT was introduced, which proved successful in
performing high loops gauge invariant computations in lattice QCD \cite{8l}. 

The idea which makes NSPT possible is the following: 
The Langevin equation provides a quantisation scheme for field theories,
and also a tool for numerical computations of the Functional Integral.  
It may be introduced by giving the fields a dependence on a new parameter
(stochastic time), $ U(\mu,x) \longrightarrow U_{\eta}(\mu,x,t),
\hskip2mm  x\in R^4$ ($\eta(t)$ is an appropriate random process), and
then imposing the fields to obey the Langevin equation, 
\beq
\frac{\partial}{\partial t} U_{\eta} = 
( -i \nabla S[U_{\eta}]_{|_{\mu,x,t}} - i \,\eta ) 
U_{\eta}
\eeq
which determines the evolution of the fields in stochastic time. 
One then notices in the Langevin equation the explicit dependence 
on $g$ through the action $S$; as a result the solution $U_{\eta}$ 
depends on $g$ as well and can be formally expanded as a 
power series in $g$.
This produces a cascade of
stochastic equations, of which only the first depends explicitily on 
the noise $\eta$ and each one depends only on the lower order fields.

\section{Symanzik's analysis}

In order to get a continuum--limit result from perturbative
calculations on  the lattice it is necessary to recognize in the data
the dependence on the cutoff. Typically one computes a coupling which is 
defined at a length scale $L$ on a lattice of size $I=L/a$, thus obtaining 
a generic perturbative coefficient $m(I)$; 
then Symanzik's analysis suggests an asymptotic expansion for $m(I)$
in the form:
$ m(I) = \sum_{n=0}^{\infty}\sum_{k=0}^{l} \frac{c_{n,k}\log(I)^k}{I^n},$
where the maximum power of the $\log$ is fixed by the order of the 
computation. 
It is actually this analysis that sets the required precision of the 
computation.

\section{NSPT for the Schroedinger Functional scheme}

One particulary interesting renormalisation scheme is introduced via 
the running coupling $\bar{g}^2(L)$  introduced in
\cite{FS} and based on the Schroedinger Functional with an induced
background field, obtained by fixing boundary conditions 
in the $t=0$ and $t=L$ time--slices.
The coupling $\bar{g}^2(L)$ is defined through the rensponse to a 
change in the boundary conditions \cite{FS}.
Through a finite size scaling technique \cite{fsst} the ALPHA Collaboration 
has been able to compute this coupling on a huge momentum range up to 
the perturbative domain in which contact can be made with other schemes 
such as $\MSbar$.
Beeing this scheme a successful one with many respects and having 
perturbative computations of this coupling already been performed at two loop 
level (so that cross-checks are possible), it was a natural candidate 
for our analysis. 

It is however the first time that NSPT is applied
starting from a vacuum which is not the trivial one. In fact the
fields have to be represented as perturbative expansion around the
classical solution induced by the background field 
$ U_{\eta} = V + \sum_{k > 0} \, \b^{-\frac{k}{2}} \, U^{(k)}_{\eta}$.
This poses no problem in principle, but it will be interesting to see
which kind of effects such a modification brings to the precision of
the results.
As far as the observable is concerned, it is known to exhibit quite strong 
fluctuations even at non--perturbative level. A feature that
immediately emerges is that it is effectively defined only on the two
boundary time slices, and not on the bulk.

As it has been soon recognized NSPT in gauge field theories is
hopeless without introducing any kind of stochastic gauge fixing (see
below).  In fact, even if this is not necessary in principle, diverging
fluctuation will soon dominate without it \cite{Zwa,RDL}.
A natural way of doing it  is to apply (after every Langevin evolution step)
a gauge trasformation $ U^w_\mu(n) \, = \, e^{w(n)} \, U_\mu(n) \,
e^{- w(n+\hat{\mu})} $ where $ w(n) =  - \frac{\alpha}{2} 
\sum_\mu \partial^L_\mu (U_\mu(n) - U^{\dag}_\mu(n))_{\mbox{{\small
traceless}}} $ and $( 0 < \alpha < 1 )$.
It is easy to verify that such a trasformation attracts the system
towards configurations where the norms of the fields are under control.
One can also apply a slightly modified version of the Landau gauge
prescription (because of the presence of a non trivial vacuum),
however this produces no significant difference in the results.

Unfortunately the results for the observable defining $\bar{g}^2(L)$
are extremely noisy, preventing any Symanzyk analysis to get a
continuum limit. Perturbative calculations had been carried on analytically in
\cite{pertan} with very high precision to the second non trivial
order. These known results are reproduced, for both $SU(2)$ and
$SU(3)$, and it would be easy to go to higher orders. However the
errorbars makes these results useless in the context they are
interesting.

Two considerations are in order: first the imprecision is due
essentially to the observable and not to the new vacuum configuration
upon which perturbations are made, nor to the different choice of
gauge fixing one is forced to take. In fact simulations of the
plaquette have the same precision we were used to. This suggests that
this approach could be more profitably applied to observables build up
with Wilson loop, leaving still the freedom of choice of other
conditions.
Secondly we convinced ourself that a great help could come from a more
efficient procedure of gauge fixing \cite{gf}. This is the subject of
what follow. 

\section{Fadeev \& Popov in NSPT}

We now propose a new formalism which is even farer from the original
spirit \cite{PW} (``Perturbation Theory without Gauge Fixing''), but implements
the Fadeev \& Popov mechanism in a (perturbative) Monte Carlo.
Such a formalism consists in the application to NSPT of a technique
which is well known in the standard Langevin equation \cite{BKKLSW}
and is in principle a solution for NSPT ``dynamical fermions'' as well.

Consider any problem of the class $e^{-S}\det M$ (fermions, F\& P). In
principle $e^{-S}\det M \sim e^{-(S-Tr\log M)}$. A solution in 
the standard Langevin equation has been known for a long time:consider
$U(t+1)= e^{-f T}U(t)$  
where $f_i = \eps [\partial_i S_g - 
\Re [\xi_k^{\dag} M_{k l}^{-1}\partial_i M_{l n} \xi_n]] $
and $\xi_k$ ($k$ multi index) are  Gaussian as well with 
$\langle |\xi|^2 \rangle=1$.
Again to  $O(\eps)$ what matters in Fokker--Plank equation is
(average on $\xi$) $ \langle f_i \rangle = \eps \partial_i [S_g - Tr
\log M] + \sqrt{\eps} \eta $
(where $\eta$ are Gaussian as well with $\langle \eta^2 \rangle=2$)
which means for the probability density of the fields $P \sim
e^{-(S-Tr \log M)} = e^{-S} \det M $

Instead of the fermion determinant consider the
functional integral which is relevant for PT in covariant gauges (i.e.
Fadeev \& Popov mechanism)  $\langle O \rangle = 1/Z \int DU e^{-(S_g+
S_{gf})} \Delta_{FP}[U] O[U] $
where $S_g =$ gauge Wilson action, $S_{gf} $ is the gauge fixing
action, and $\Delta_{FP}[U] $ the Fadeev Popov determinant.
$S_{gf}$ is in fact the action in charge of ``killing the
divergences'' in Langevin equation. 

It is necessary to compute the contributions to the equation of motion
coming both from $S_{gf}$ and from $\Delta_{FP}$
The first gives no problem. As far as the second is concerned
notice that of course nobody implements $\Delta_{FP}$ via
$\Delta_{FP} = \det L = e^{Tr \log L}$. The resulting action would
be strongly non local.
Instead $\det L$ is in standard applications obtained by a functional
integral over grassmann fields (ghosts).
Shadow of the non locality of the would be $S_{FP}[U]$ is $L^{-1}$ in 
$ f_i = \eps [\partial_i (S_g+ S_{gf}) - \Re [\xi^{\dag}
L^{-1}\partial_i L \xi]]$ 
but the crucial observation is now that
$ L^{-1} = L_0^{-1} - g L_0^{-1} L_1 L_0^{-1} + 
g^2 (L_0^{-1}L_1L_0^{-1}L_1L_0^{-1} - L_0^{-1}L_2L_0^{-1}) + O(g^3)$.
The message is that  $L^{-1}$ has a simple  recursive form easy to
guess also at this order; only $L_0^{-1}$ is a proper inverse,
which is well known and easy in Fourier space and it does not depend
on the fields. 

\section{Conclusions}

NSPT proved successful in performing high loop computation in lattice
$QCD$. The precision demostrated in the computation of different
quantities show that the task of pushing one loop further the
evaluation of $SU(3)$ ($SU(2)$) $\be$ function is attainable. 

However the choice of the scheme based on the Schroedinger Functional
(which has many desirable features) proved unlucky for our
approach. The  direction to pursue is now double. On one side one can
examine different schemes of renormalization (for instance based on
Wilson loops), where NSPT calculations are more precise, and determine
in this way the next coefficient of the $\be$ function. 
On the other side one can look at the possibility of implementing the
Fadeev \& Popov gauge fixing in NSPT. 
One notes that  contributions to $L^{-1}$ are out of a recursive
structure and depend only on one inverse $L_0^{-1}$, which is easy to
manage in Fourier space. 
$L_0^{-1}$ does not depend on the fields.
Because of the former observation, the actual implementation of
the proposed method depends strongly on the architecture of the
computer: without a good FFT it is hopeless.
 Same sort of remarks apply to including dynamical fermions
contributions as well.

\end{document}